\documentclass[twocolumn,times,tighten]{aastex631}
\usepackage{amsmath,amstext}
\usepackage[english]{babel}
\usepackage[utf8x]{inputenc}
\usepackage{textgreek}

\newcommand{\teff}{$T_\mathrm{eff}$}

\newcommand{\logg}{$\log g$}
\newcommand{\feh}{[Fe/H]}
\newcommand{\micro}{$\xi_\mathrm{micro}$}

\newcommand{\kms}{km\,s$^{-1}$}
\newcommand{\mic}{$\mu \mathrm m$}

\begin{document}

\title{Chemical Abundances in the Nuclear Star Cluster of the Milky Way:\\
$\alpha$-Element Trends and Their Similarities with the Inner Bulge}

\correspondingauthor{Nils Ryde}
\email{nils.ryde@fysik.lu.se}

\author[0000-0001-6294-3790]{Nils Ryde}
\affil{Division of Astrophysics, Department of Physics, Lund University, Box 118, SE-22100 Lund, Sweden}

\author[0000-0002-6077-2059]{Govind Nandakumar}
\affil{Division of Astrophysics, Department of Physics, Lund University, Box 118, SE-22100 Lund, Sweden}
\affil{Aryabhatta Research Institute of Observational Sciences, Manora Peak, Nainital 263002, India}

\author[0000-0002-6590-1657]{Mathias Schultheis}
\affil{Université Côte d’Azur, Observatoire de la Côte d’Azur, Laboratoire Lagrange, CNRS, Blvd de l’Observatoire, 06304 Nice, France}

\author[0000-0002-9035-3920]{Georges Kordopatis}
\affil{Université Côte d’Azur, Observatoire de la Côte d’Azur, Laboratoire Lagrange, CNRS, Blvd de l’Observatoire, 06304 Nice, France}

\author{Paola di Matteo}
\affil{Observatoire de Paris, section de Meudon, GEPI, 5 Place Jules Jannsen, 92195 Meudon, France}

 \author{Misha Haywood}
  \affil{Observatoire de Paris, section de Meudon, GEPI, 5 Place Jules Jannsen, 92195 Meudon, France}

\author[0000-0001-5404-797X]{Rainer Schödel}
 \affil{Instituto de Astrofísica de Andalucía (CSIC), Glorieta de la Astronomía s/n, 18008 Granada, Spain}

 \author[0000-0002-6379-7593]{Francisco Nogueras-Lara}
 \affil{European Southern Observatory, Karl-Schwarzschild-Strasse 2, 85748 Garching bei München, Germany}

 \author[0000-0003-0427-8387]{R. Michael Rich}
 \affil{Department of Physics and Astronomy, UCLA, 430 Portola Plaza, Box 951547, Los Angeles, CA 90095-1547, USA}

\author[0000-0002-5633-4400]{Brian Thorsbro} 
 \affil{Université Côte d’Azur, Observatoire de la Côte d’Azur, Laboratoire Lagrange, CNRS, Blvd de l’Observatoire, 06304 Nice, France}

 \author[0000-0001-7875-6391]{Gregory Mace}  
 \affil{Department of Astronomy and McDonald Observatory, The University of Texas, Austin, TX 78712, USA}

\author[0000-0002-4287-1088]{Oscar Agertz}
    \affil{Division of Astrophysics, Department of Physics, Lund University, Box 118, SE-22100 Lund, Sweden}

\author[0000-0002-3181-3413]{Anish M. Amarsi}
 \affil{Theoretical Astrophysics, Department of Physics and Astronomy, Uppsala University, Box 516, SE-751 20 Uppsala, Sweden}

 \author{Jessica Kocher}
 \affil{Division of Astrophysics, Department of Physics, Lund University, Box 118, SE-22100 Lund, Sweden}

 \author{Marta Molero}
 \affil{Institut für Kernphysik, Technische Universität Darmstadt, Schlossgartenstr. 2, Darmstadt 64289, Germany}

 \author[0000-0002-6040-5849]{Livia Orglia}
 \affil{INAF, Osservatorio di Astrofisica e Scienza dello Spazio di Bologna, Via Gobetti 93/3, 40129 Bologna, Italy}

 \author{Giulia Pagnini}
 \affil{GEPI, Observatoire de Paris, PSL Research University, CNRS, Place Jules Janssen, F-92195 Meudon, France}

 \author[0000-0001-9715-5727]{Emanuele Spitoni}
 \affil{INAF, Osservatorio Astronomico di Trieste, Via Tiepolo 11, 34131 Trieste, Italy}

\begin{abstract}

A chemical characterization of the Galactic Center is essential for understanding its formation and structural evolution. Trends of $\alpha$-elements, such as Mg, Si, and Ca, serve as powerful diagnostic tools, offering insights into star-formation rates and gas-infall history. However, high extinction has previously hindered such studies. In this study, we present a detailed chemical abundance analysis of M giants in the Milky Way’s Nuclear Star Cluster (NSC), focusing on $\alpha$-element trends with metallicity. High-resolution, near-infrared spectra were obtained using the IGRINS spectrograph on the Gemini South telescope for nine M giants. Careful selection of spectral lines, based on a solar-neighborhood control sample of 50 M giants, was implemented to minimize systematic uncertainties. Our findings show enhanced $\alpha$-element abundances in the predominantly metal-rich NSC stars, consistent with trends in the inner bulge. The NSC stars follow the high-[$\alpha$/Fe] envelope seen in the solar vicinity’s metal-rich population, indicating a high star-formation rate. The $\alpha$-element trends decrease with increasing metallicity, also at the highest metallicities. Our results suggest the NSC population likely shares a similar evolutionary history with the inner bulge, challenging the idea of a recent dominant star formation burst. 
This connection between the NSC and the inner-disk sequence suggests that the chemical properties of extragalactic NSCs of Milky Way type galaxies could serve as a proxy for understanding the host galaxies' evolutionary processes.

\end{abstract}


\section{Introduction}
\label{sec:intro}

In recent years, considerable effort has been devoted to reach a holistic picture of  the Galactic Center region of the Milky Way.  
The goal is to get a better understanding of the relation between its formation history and that of the Milky Way as a whole, but also  a better understanding of the roles played by galactic nuclei in galaxy formation and evolution, in general \citep[see, e.g,][]{neumayer:20,jwst_nsc_white:23}. A chemical characterization of this region is a critical component of this endeavor.

The center of the Milky Way consists of the Nuclear Star Cluster (NSC), the Nuclear Stellar Disk (NSD), and the Central Molecular Zone (CMZ).
The NSC is a compact, spherical stellar structure with a half-light radius of $\sim4$\,pc, a diameter of $\sim$\,12 pc ($\sim300\arcsec$),  and a total stellar mass of $2\times 10^7\,\mathrm M_\odot$ \citep[e.g.][]{schodel:14a,schodel:14b,feld:14,chat:15,fritz:16,feld:17,gallego:20}. The surrounding NSD consists of a flat, rotating, disk-like structure with a break radius of $\sim90$\,pc, a vertical height of $\sim45$\,pc, and a total stellar mass of  $1\times 10^9\,\mathrm M_\odot$ \citep[see, e.g.,][]{launhardt:02,bland:16}. The NSD lies within the Central Molecular Zone (CMZ), which is $\sim 250\times50$\,pc large \citep{henshaw:23,jwst_nsc_white:23} and  the Galactic Center region is, as a whole,  embedded in the Galactic Bulge/Bar \citep{sormani:22}.

Large gas inflows channeled by the Galactic bar have made the Galactic Center region the largest reservoir of dense gas in the Milky Way \citep{baba:20,sormani:22}. Populations of young stars provide evidence for recent star formation in the predominantly old stellar populations with ages between 8-10\,Gyrs \citep{matsunaga:11,lara:20,schodel:20,thorsbro:23,gallego:24}.  
The star-formation histories (SFH) have varied with time and appear different in the NSD and NSC \citep{schodel:20,lara:21}.  
While there are differences in the stellar populations and formation histories of the NSD and NSC, \citet{lara:23} suggest,  based on kinematics and metallicity gradients of the two systems, a smooth transition between the two and that they might be parts of the same structure. In the NSC, the majority of star formation was determined by \citet{schodel:20} to have occurred more than 10 Gyrs ago. However, it also seems to contain an intermediate-aged population, approximately 3 Gyrs old, which accounts for less than 15\% of the total star formation \citep{schodel:20,lara:21}. In contrast, \citet{chen:23} identify a dominant, metal-rich component (constituting over 90\% of the stellar mass) in the NSC, with an age of around 5 Gyrs. 

There are two primary formation scenarios proposed for the NSC \citep{neumayer:20}: {\it (i)} In-situ formation, where gas is funneled into the central few parsecs, triggering star formation. This process may be driven by several mechanisms, including bar-driven gas infall, dissipative nucleation, tidal compression, or magneto-rotational instability (see \citealt{neumayer:20} for further discussion). (ii) Infall of massive stellar clusters into the galactic nucleus, which could account for the metal-poor population observed in the NSC \citep[e.g.,][]{capu:93,mastro:12,Hartmann2011, Arca-Sedda2014}. Indeed, recent work using cosmological zoom simulations \citep[][]{Gray2024} has demonstrated how metal-poor ([Fe/H]$\lesssim -1$ ) NSCs can form in starbursting dwarf galaxies at high redshifts, with cluster properties in agreement with observations. However, globular cluster infall cannot explain the presence of young stars \citep[e.g.,][]{Feldmeier-Krause2015} and globular clusters observed in the Milky Way today cannot have contributed a significant fraction of the mass of old stars in the NSC \citep{Dong:2017}. It is most likely that the NSC formed through a combination of both these scenarios.

Abundance trends versus metallicities for the different stellar populations in the Galactic Center can add new pieces of evidence. Such studies would complement investigations of dynamics, kinematics, star-formation histories, and the distribution of metallicities. Since different elements represent different nucleosynthetic channels with their own evolutionary timescales \citep[see, e.g.,][]{Matteucci:2021,manea:23}, the trends will be different depending on, for instance, the star-formation rate and gas infall, and can, therefore, reveal differences between stellar populations.

Detailed abundance measurements near the plane and Galactic Centre, however, still remain relatively sparse.   Due to high dust-extinction \citep{nishiyama:06,lara:18}, high-resolution spectroscopic observations of stars in the Galactic Center need to be carried out in the near-infrared (H and K bands) regime.  
Abundance trends of the Galacitc Center region   
first appeared in \citet{cunha:07}. They focused on the young supergiant population in the NSC showing elevated $\alpha$-elements at high metallicities. Later, \citet{ryde:15,ryde:2016_metalpoor,ryde:2016_bp2,Nandakumar:18,Nieuwmunster:2023} presented $\alpha$-element trends for 9 giants from the old 
population probably in the NSD and not in the NSC ($2.5-5.5$\,\arcmin\ North 
of the very center, which corresponds to a projected galactocentric distance of $5-10$\,pc) showing more thick-disk-like behaviour for the metal-rich stars there. A few years later, \citet{do:18} derived elevated abundances of V, Sc, Y from strong spectral lines in two cool, metal-rich stars within 0.5 pc from the very center, i.e. in the NSC. These strong lines were instead explained in \citet{thorsbro:2018} as a line-formation property in cool giants, rather than being due to a high abundance. In the first [Si/Fe] determinations for old stars actually in the NSC, \citet{thorsbro:2020} find a disk-like trend at subsolar metallicities, but an enhanced trend at supersolar metallicities. From a dynamical investigation, 
15 of the stars investigated were determined to be located in the NSC and 5 in the NSD. The near-infrared spectroscopic survey  APOGEE \citep[Apache Point Observatory Galactic Evolution Experiment;][]{Holtzman:2018,Jonsson:2018}  
also targeted these inner regions, although only a handful of their stars lie in the projected region of the NSD\footnote{The fibre diameter of $2\arcsec$ used in APOGEE is not adequate to study the Galactic Centre given the high stellar crowding.}. They determined the $\alpha$-abundance trends for 
stars in their GALCEN field, centered at $(l,b)  \sim 
(+0^{\circ}.2,-0^{\circ}.1)$, showing trends similar to that observed in the Galactic bulge
\citep{Schultheis:2015,Schultheis:2020}.

The aim of the  present study is to investigate the abundance trends of the $\alpha$ elements Mg, Si, and Ca as a function of metallicity for a sample of stars 
in the Nuclear Star Cluster, differentially against the trends from a control sample of the same type of stars in the solar neighbourhood \citep[based on the study by][]{Nandakumar:2023}. We will thus compare detailed abundances of M giants in both stellar populations, observed with the highest resolution infrared spectrograph available and analysed in the same way, which minimizes any possible systematic uncertainties. We search for differences that may reflect a different chemical evolution in the NSC, compared to the solar vicinity.

 \begin{figure}
  \includegraphics[trim={0 4cm 0 0},clip,angle=-90,width=0.55\textwidth]{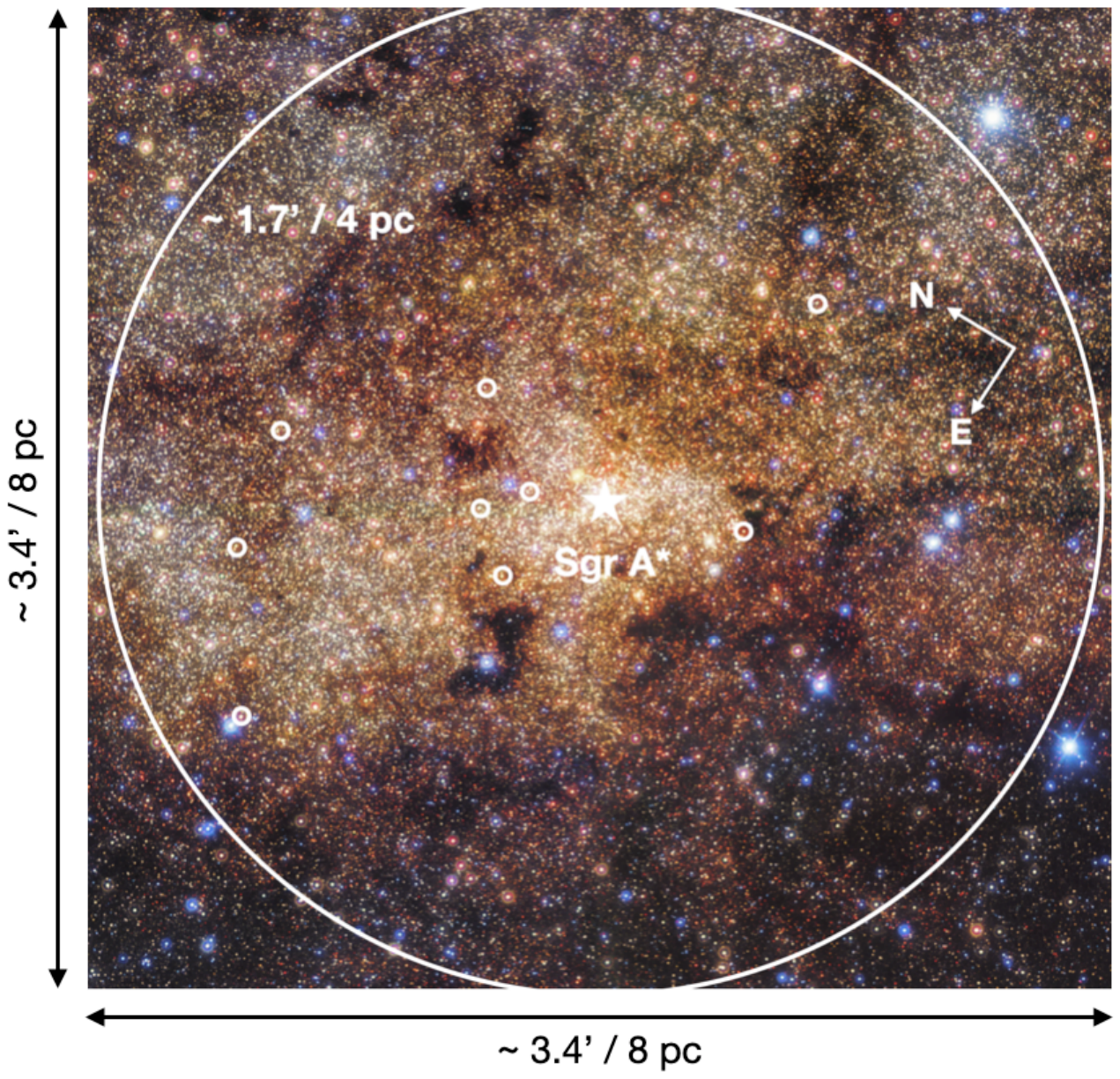}
  \caption{GALACTICNUCLEUS JHK$_s$ image of the Milky Way's NSC \citep{lara:18,lara:19} showing the 9 stars for which we have obtained spectra, marked by white small circles.  The image scale is 8x8 pc$^2$ and the large white circle indicates the effective radius of the NSC ($\sim 4$\,pc, or $\sim 100\arcsec$), while the position of the supermassive black hole, A$^*$, is denoted by a white star in the middle of the image.
The directions East (E) and North (N) are labeled in the upper right corner. }
  \label{fig:map}
\end{figure}

This paper is the third in a series characterizing the chemistry in the intermediate-age to old red giants populations of the Galactic Center with high-resolution, near-IR spectroscopy. The first paper \citep{Rich:2017}, which was based on KECK/NIRSPEC observations of 15 M giants in the NSC, found a broad metallicity distribution ranging from $-0.5<$\feh$<+0.5$\,dex, with most of the stars being at or below the solar iron abundance. This is broadly similar to other fields in the Galactic bulge, and not as narrow as that found for supergiants in the NSC \citep{cunha:07}. In the second paper \citep{thorsbro:2018}, we  explore the silicon abundance trends versus metallicity for these 15 NSC stars. The trends of the $\alpha$ elements are diagnostically interesting and are sensitive to the star-formation rates in the stellar population. We find a thick-disk trend for subsolar metalicities, but a surprising [Si/Fe] enhancement for supersolar metallicities  of $+0.25<$\feh$<+0.5$ in 4 NSC stars and 2 NSD stars. To confirm this result, it is necessary to study the trends of more $\alpha$ elements in a bigger sample of stars. We note again that \citet{cunha:07}  found enhanced alpha elements in the supergiant population, while \citet{ryde:15,Nieuwmunster:2023}  found trends for Mg, Si, and Ca resembling the enhanced disk trend for NSD stars. Similar results were found by \citet{Schultheis:2020} for the NSD.

In this study we will, thus,  present the detailed $\alpha$ abundances of the Nuclear Star Cluster observed at a spectral resolution that resolves stellar lines. In a forthcoming paper in the series, we will further discuss elemental abundance trends for other  groups of elements such as fluorine, odd-Z, iron-peak, and neutron-capture elements. The details of the observations and data reduction are provided in Section 2, followed by the details of the spectroscopic analysis in Section 3. In Section 4, we show the elemental abundance trends together with the identically analaysed 50 solar neighborhood stars. We discuss our results  in Section 5 and conclude in Section 6.

\section{Observations and Data Reduction}
\label{sec:obs}

\begin{deluxetable*}{l r r r r r r r r r}
\tabletypesize{ }
\tablewidth{0pt}
\tablecaption{Observational details of M giant stars. \label{table:obs}}
\tablehead{
 \colhead{Name\tablenotemark{a}}  & \colhead{RA} & \colhead{DEC} & \colhead{H$_\mathrm{}$\tablenotemark{b}} & \colhead{K$_\mathrm{s}$\tablenotemark{b}}  & \colhead{Date}  & \colhead{Exp. Time}   & \colhead{S/N$_\mathrm{H}$ / S/N$_\mathrm{K}$\tablenotemark{c}} &  \colhead{id\tablenotemark{d}} & Telluric star\\
 \colhead{} & \colhead{h:m:s} & \colhead{d:m:s} & \colhead{[mag]} & \colhead{[mag]} &  \colhead{UT}  & \colhead{[s]}  &  \colhead{per res. element} & & A0V} 
\startdata 
FK48$^{(2)}$ & 17:45:41.301 & -29:00:08.406 &  12.53  & 10.76 &  2023-04-30   &  
2240 & 35/105    & 48 & HIP86098 \\
FK5020265$^{(2)}$ & 17:45:46.187 & -28:59:48.253  &  11.84  & 9.91 &  2023-03-24   &  
840 & 50/150    & 5020265 & HIP86098  \\
FK87$^{(1)}$ & 17:45:40.671 & -29:00:15.318  &  13.04  & 10.75 &  2022-05-16   &  
1620 & 45/175   & 87 & HIP86098  \\
Feld31$^{(2)}$ & 17:45:42.000 & -29:00:20.000  &13.38 & 10.61  &  2023-04-26   &  
1200 & 20/150   & 31  & HIP86552\\
Feld84$^{(2)}$ & 17:45:39.400 & -29:00:58.900  &13.90 & 10.66  &  2023-04-28   &  
2080 & 10/135   & 84 & HIP86098 \\
GC15540$^{(2)}$ & 17:45:41.900 & -28:59:23.390  &  12.57  & 10.49  &  2023-03-24   &  
2200 & 35/120  & 3001047 & HIP86098 \\
GC16890$^{(2)}$ & 17:45:43.900 & -28:59:28.500  &13.23 & 10.72  &  2023-04-26   &   
2240 & 20/105   & 3000179  & HIP88152 \\
GC13727$^{(3)}$ &  17:45:39.590 & -28:59:56.210 & 13.11 & 10.82  &  2024-04-11   & 
2520 & 20/120   & 73 & HIP94663 \\
GC16895$^{(3)}$ & 17:45:35.640 & -29:00:47.000  & 13.16& 10.75  &  2024-04-20   & 
2400 & 20/115   & 1012095 & HIP86098 \\
\enddata
\tablenotetext{a}{Gemini-S Progamme identification:
$^{(1)}$  GS2022A-Q-208;  $^{(2)}$ GS2023A-Q-304; $^{(3)}$ GS2024A-Q-304.} \tablenotetext{b}{The H  and K magnitudes are from \citet{Nishiyama2013}} \tablenotetext{c}{The signal-to-noise ratios were provided by RRISA \citep[The Raw $\&$ Reduced IGRINS Spectral Archive;][]{rrisa} and are the average S/N for the H- and K-bands per resolution element. The S/N varies over the orders and is lowest at the ends of the orders.} \tablenotetext{d}{Identification number from \citet{Feldmeier-Krause:2017,Feldmeier-Krause:2020}. }
\end{deluxetable*}
\vspace{-0.83cm}

We have determined the stellar parameters and the $\alpha$ elements for 9 M-giants (\teff$< 4000$~K) in the NSC from high-resolution spectra observed with the Immersion GRating INfrared Spectrograph \citep[IGRINS;][]{Yuk:2010, Wang:2010, Gully:2012, Moon:2012, Park:2014, Jeong:2014}. IGRINS provides spectra with a spectral resolving power of $R \sim$ 45,000, spanning the full H and K bands (1.45 - 2.5 $\mu$m), 
thus giving access to a wealth of spectral lines enabling a detailed study of stellar abundances for a range of different elements.

The stars were observed with IGRINS mounted on the Gemini South telescope \citep{Mace:2018} under the programs 
GS-2022A-Q-208, GS-2023A-Q304, and GS-2024A-Q-304. These were observed in service mode over a period from May 2022 to April 2024 and their location on the sky is shown in Figure \ref{fig:map}. The stars are presented in Table \ref{table:obs}, where the H and K magnitudes are given, as well as the date of observation, exposure times, and achieved signal-to-noise ratio (S/N) per resolution element in the H and K bands, respectively. The Gemini Program identifications for the three runs are also provided for every star.

The IGRINS observations were conducted using one ABBA nod sequence (for the 2022, and 2023 runs) and two nod sequences (for the 2024 run)  along the slit to facilitate sky background subtraction. Exposure times were chosen to achieve an average S/N of approximately 100 in the K band, resulting in observing times ranging up to 50 minutes. Despite the higher sensitivity of the IGRINS H-band detector, the significant reddening affects the H band more than the K band \citep{lara:20}, resulting in lower S/N in the H band.

The IGRINS data were 
reduced with the IGRINS PipeLine Package \citep[IGRINS PLP;][]{Lee:2017} to optimally extract the wavelength calibrated spectra, order by order, after flat-field correction and A-B frame subtraction \citep{Han:2012,Oh:2014}. The H-band spectra range from $1.45-1.83$\,\mic\ over 27 orders, and the K-band from $1.87-2.49$\,\mic\ over 26 orders. The wavelength solution is based on sky OH emission lines.

The contaminating telluric lines were reduced by dividing the target spectra with a telluric standard-star spectrum, showing no stellar features, in our case fast-rotating late-B to early-A dwarfs. These were observed close in time and at an air mass similar to that of the science targets. In the subsequent abundance analysis, we keep track of the spectral regions impacted by, but corrected from telluric lines 
ensuring that any possible residuals resulting from the correction process are taken care of.

The spectral orders of the science targets and the telluric standards were subsequently stitched together after normalizing every order and then combining them in {\tt iraf} \citep{IRAF}, excluding the low S/N edges of every order. This resulted in one normalized stitched spectrum for the entire H and K bands. The spectra were then resampled and to take care of any residual modulations in the continuum levels, we  define specific local continua around spectral lines being studied. Finally, the spectra are shifted to laboratory wavelengths in air after a stellar radial velocity correction.  We also allowed for small wavelength shifts in the abundance analysis of every spectral line, which allows any errors or trends in the wavelength solution to be taken into account. 

Especially toward the Galactic Center, with the large extinction in the line-of-sight, broad absorption features caused by DIBs \citep[diffuse interstellar bands; see, e.g.,][]{dib:geballe1}  might appear in high-resolution spectra. These are probably due to large molecules in the interstellar medium (ISM). Most DIBs have been identified in optical spectra, but some are found in the NIR \citep{dib:geballe_nature}, and more are being identified with new instruments 
\citep[e.g., with X-shooter, APOGEE, IGRINS, WINERED, and CRIRES in][respectively]{dib:cox,dib:elya,dib:galazut,dib:winered,dib_crires}. None of these known DIBs are, however,  close to the spectral lines that we use for our abundance determination in this work.

\begin{deluxetable*}{c r c r c r r r r r r r  r }
\tabletypesize{ }
\tablewidth{0pt}
\tablecaption{Stellar and orbit parameters, and C, N, and O abundances for the NSC giants.} \label{table:parameters}
\tablehead{
 \colhead{Star}  & \colhead{$T_\mathrm{eff}$} & \colhead{$\log g$}  & \colhead{[Fe/H]}  &  \colhead{$\xi_\mathrm{micro}$} & \colhead{[C/Fe]} & \colhead{[N/Fe]} & \colhead{[O/Fe]\tablenotemark{a}} & \colhead{$\rm \mu_{l}$}& \colhead{$\rm \mu_{b}$} & \colhead{$v_\mathrm{los}$} & \colhead{$r_\mathrm{apo}$\tablenotemark{b}} &\colhead{$z_\mathrm{max}$} \\
& \colhead{[K]} & \colhead{[$\log$(cm\,s$^{-2}$)]} & \colhead{[dex]} & \colhead{[\kms]} & \colhead{[dex]} & \colhead{[dex]} & \colhead{[dex]} & \colhead{[mas/yr]}   &  \colhead{[mas/yr]} & \colhead{[\kms]} & \colhead{[pc]} & \colhead{[pc]}  }
\startdata 
FK48  &  3440  &  0.6  &  0.12  &  1.8  &  0.12  &  0.25  &  0.09  & -2.54 & -0.42& 0 & 9.81 & 6.70\\ 
FK5020265  &  3350  &  0.6  &  0.26  &  1.9  &  0.06  &  0.24  &  0.04  &-3.94 & 5.71&157 & 9.93 & 8.47\\ 
FK87  &  3352  &  0.5  &  0.09  &  1.9  &  0.07  &  0.35  &  0.11  &-- &-- &  196 & -- & --\\ 
Feld31  &  3825  &  1.5  &  0.42  &  2.3  &  -0.04  &  0.5  &  -0.04& -1.02& 3.25 & 118 &10.32 & 7.29 \\ 
Feld84  &  3709  &  1.4  &  0.48  &  2.7  &  0.03  &  0.47  &  -0.07  &-- &-- & 0 & -- &--\\ 
GC15540  &  3361  &  0.5  &  0.11  &  1.9  &  0.03  &  0.36  &  0.10  &-2.68& 2.57 & 0 & 8.17 & 6.21 \\ 
GC16890  &  3423  &  0.7  &  0.19  &  2.0  &  0.02  &  0.33  &  0.06  & -2.07&-1.30& 236 & 8.85 & 7.02\\ 
GC13727  &  3356  &  0.4  &   0.00  &  2.1  &  0.20  &  0.32  &  0.16  & 0.74 & -0.97& -59 &9.17 & 6.76\\ 
GC16895  &  3366  &  0.4  &  -0.12  &  2.2  &  0.19  &  0.45  &  0.21  & -7.93 &0.33& -39 & 11.10 & 5.03 \\
\enddata
\tablenotetext{a}{The [O/Fe] is provided from a simple functional form of the [O/Fe] versus [Fe/H] trend in \citet{Nandakumar:2023} based on the \cite{Amarsi:2019} trend.}
\tablenotemark{b}{$r_{apo}$  is the maximum apocentric radius from the galactic center projected on the x-y plane. }
\end{deluxetable*}
\vspace{-0.83cm}

\section{Analysis}
\label{sec:analysis}

We determine the abundances of the $\alpha$-elements Mg, Si, and Ca for 9 M giants in the NSC from their high-resolution spectra and compare the trends of these abundances with those determined for 50 M giants in the solar neighbourhood. 

\subsection{Nuclear Star Cluster Targets}

A detailed abundance analysis of the stellar populations in the NSC requires the use of red giants as probes, observed at high spectral resolution. The significant optical extinction along the line of sight to this region precludes optical analyses. However, reliable abundance determinations can be achieved using high-resolution infrared spectra of K and M giants. In practice, the stars in the NSC that are sufficiently bright for high spectral resolution observations with 10-meter class telescopes are primarily M giants, with effective temperatures below 4000 K. This is attributed to the brightness of these stars, their spectral energy distributions, and the metallicity distribution within the inner regions of the Milky Way. Note that we have deliberately excluded the young M-type supergiants in the NSC from our analysis \citep[such as IRS 7 discussed by][]{carr:2000,irs7:22}\footnote{In this study, we differentially determine abundances for low-mass M giants, using a reference sample of the same stellar type. M supergiants, being much more massive, require their own specialized spectroscopic analysis. Our focus here is on M giants, while the young stellar populations in the NSC will offer a complementary perspective to the findings presented in this work \citep[see, e.g.,][]{thorsbro:23}}. 
The analysis has been limited to M giants with effective temperatures above 3350 K, for which our analysis method has been tested and validated. The targets are given in Table \ref{table:obs} including their coordinates.

In the Table we also provide the star identification from the integral-field, low-resolution ($R=4,200$) studies of stars in the NSC of \citet{Feldmeier-Krause:2017,Feldmeier-Krause:2020}. These authors argued, based on interstellar extinction, that these stars are located in the NSC and are not foreground stars.  To further ensure that our stars are bound to the NSC, we also used the dynamical properties of the stars to constrain their membership. The radial velocities are calculated by measuring the wavelength shift of well-identified spectral lines in the observed spectra relative to their corresponding laboratory wavelengths. Proper motions have  been obtained for 7 of our targets from the preliminary VIRAC1 (\citealt{Smith2018}) photometric and astrometric  reduction of the VVV ({\it VISTA Variables in the Vía L\'actea}) data \citep{Minniti2010}. Table \ref{table:parameters} gives these proper motions and the line-of-sight velocities.
We use the software package AGAMA (\citealt{Vasiliev2019}) to determine the orbital parameters by constructing a non-axisymmetric potential by combining the main components of the inner Galaxy that influence the dynamics of the stars: the inner bulge/bar, the NSD, and the NSC. We assume that our stars are located at 8.2\,kpc from the Sun.  We refer here to \citet{Nieuwmunster2024} for more details. All our targets have typical orbits confined to the NSC and were therefore most-likely formed in-situ. 

\begin{figure}
 \includegraphics[width=0.49\textwidth]{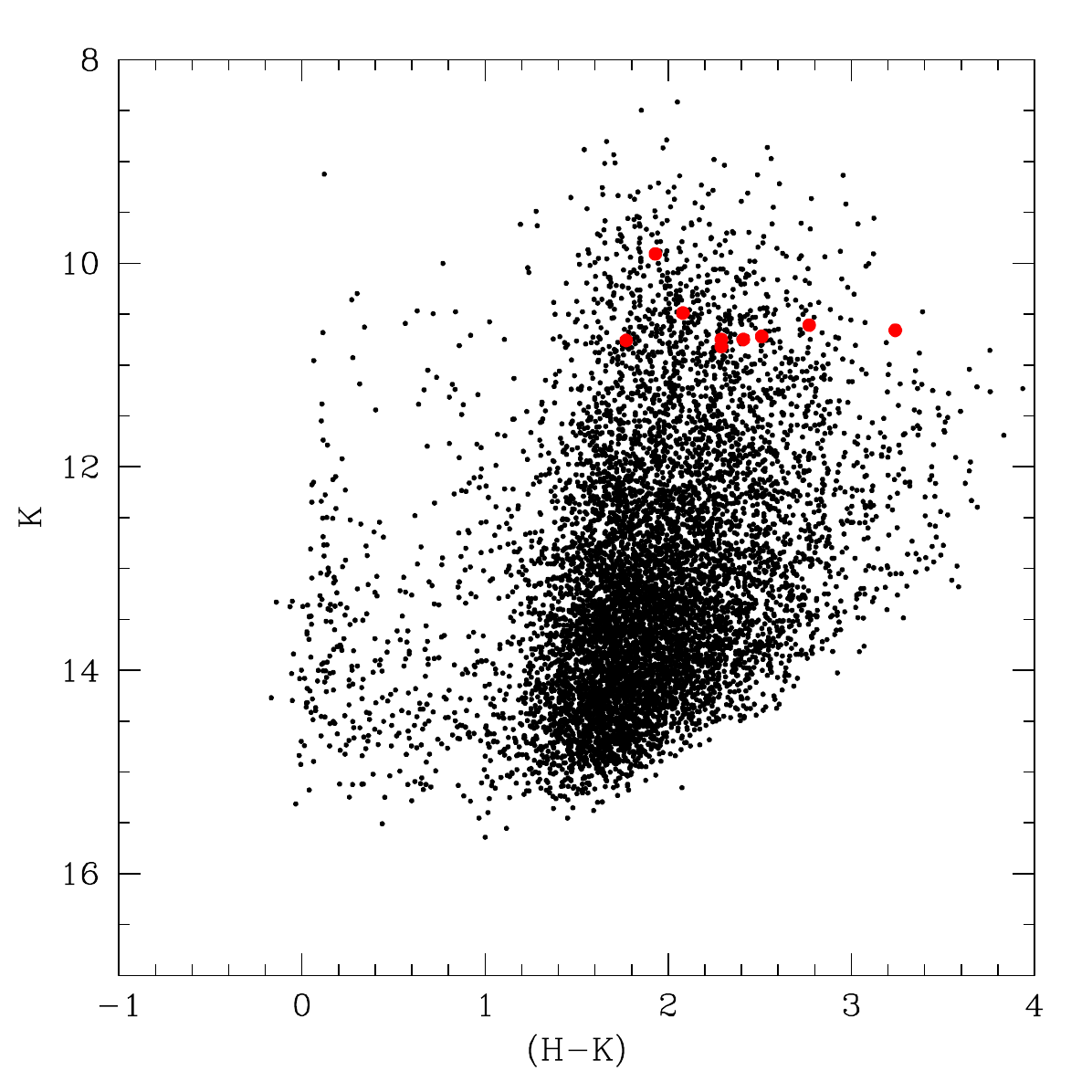}
  \caption{Colour-magnitude diagram, K versus H-K, of the NSC (\citealt{Nishiyama2013}) with our observed targets indicated by red circles.} \label{fig:CMD}
\end{figure}

Figure~\ref{fig:CMD} shows the colour-magnitude diagram of our 9 NSC stars overlaid by the overall population in the NSC. They  cover the upper branch of the RGB and a wide (H--K) range.

\citet{lara:21c,lara:22b} demonstrated that the membership of stars in the NSC can be determined based on their interstellar reddening. Although interstellar reddening is patchy, they show that in statistical way stars in the NSC do show higher extinction and therefore redder colours in the $\rm J-K$ or $\rm H-K$ diagram. In their Figure 2, they show that stars belonging to the NSC  have $\rm H-K$ colours redder than  $\rm H -K \sim 1.8$ compared to stars in the NSD or the Galactic bulge. Our sample of stars (see Table~\ref{table:obs}) have typical colours of $\rm H-K > 1.8$.  An additional extinction measurement comes from the diffuse interstellar band (DIB) clearly visible at $\rm \lambda \sim 1.527\,\mu m$ in our spectra. Although the carrier of the DIBs are still unknown, they correlate well with interstellar reddening (see equation 3 of \citealt{Zasowski2015}).  A very rough estimate of the equivalent width of the DIB give us typical values in the range of $A_{\rm V} \sim \rm 20-25$, indeed typical for the NSC region.

Together with the above mentioned orbital properties of our stars, we are  therefore confident that all our stars belong to the NSC

\subsection{Spectroscopic Method}

The spectroscopic analysis in this work is done using the spectral synthesis method, where the stellar parameters and elemental abundances of a star are determined by fitting its observed spectrum with a synthetic spectrum. The synthetic spectrum is generated using the Spectroscopy Made Easy (SME) tool \citep[SME;][]{sme,sme_code}, which calculates the spherical radiative transfer through a relevant stellar atmosphere model defined by its fundamental stellar parameters. The stellar atmosphere model is selected by interpolating within a grid of one-dimensional (1D) Model Atmospheres in a Radiative and Convective Scheme (MARCS) stellar atmosphere models \citep{marcs:08}.

The spectral analysis of M giants has received less attention than that of K giants, as K giants are easier to model in the commonly used optical wavelength regions. As a result, the knowledge of line strengths and the identification of suitable lines in infrared spectra of M giants remain limited. To address this, an investigation of the spectra of well-studied stars is necessary to determine the most appropriate spectral lines for analysis. Systematic effects, such as unknown blending lines, can impact certain spectral lines in specific regions of the stellar parameter space. Such investigations were conducted by \citet{Nandakumar:2023,Nandakumar:24_21elements}, who provided a recommended set of lines for spectroscopic studies of M giants.  We use these recommendations in the following analysis.

\subsection{Stellar Parameters}
\label{sec:stellarparameters}

Fundamental stellar parameters - namely, the effective temperature (\teff), surface gravity (\logg), metallicity (\feh) and microturbulence ($\xi_\mathrm{micro}$) - are essential for spectroscopic analysis. These parameters provide the foundation for accurately determining elemental abundances from the respective absorption lines in the observed spectrum. Therefore, it is crucial to determine accurate fundamental stellar parameters. The stellar parameters in this work were determined using the method outlined in \citet{Nandakumar:2023} for K and M giants (\teff\,$<$ 4500 K). 

\subsubsection{Effective Temperature, Metallicity, and Microturbulence}

In this method, the effective temperature (\teff) is primarily constrained by selected OH lines sensitive to temperature, while the metallicity is determined using selected Fe lines. Since the OH lines are not only sensitive to the temperature, but also to the oxygen abundance, the main assumption in the method of \citet{Nandakumar:2023} is the value of oxygen abundance. This is taken from a simple functional form of the [O/Fe] versus [Fe/H] trend in \citet{Amarsi:2019}  for thick disk stars (see also Figure 1 in \citet{Nandakumar:2023} and the study of \citet{Rojas-Arriagada:2017}). Thus a  weakness of the method is the assumption of the oxygen trend versus metallicity. Typically, a thick- or thin-disk trend is assumed in the method. This assumption is then evaluated against the derived $\alpha$-abundance trends. A star incorrectly identified as belonging to the thin or thick disk can be detected in this way and a new corrected iteration is performed assuming a correct oxygen trend \citep[for more details, see][]{Nandakumar:2023}. A lowering of the assumed oxygen abundance by $\sim 0.2$\,dex for our NSC stars is found to result to systematically lower temperatures of $\sim 50-100$\,K. While this change is within the uncertainties, it is a systematic shift, and leads to an increase of the $\alpha$-element abundances of 0.05-0.10 dex. 

The method presented in \citet{Nandakumar:2023} was validated for the temperature range $3400 \lesssim T_{\mathrm{eff}} \lesssim 4000$ K, with the lower limit determined by the temperature of the coolest benchmark star used in their study. However, there is no indication that this represents a fundamental limit for the method; it should remain effective at lower temperatures, given the smooth variation of diagnostic lines (OH, CN, CO, and Fe) with decreasing temperature. Our sample includes stars with temperatures approximately 50 K lower, which should not pose any issues for the method. As an initial validation, we do not observe any systematic trends with temperature among our coolest stars. At lower temperatures, water lines become increasingly prominent in the spectra. To accurately model these spectral features, a complete and precise water line list is required. This line list must be thoroughly tested and validated to ensure it provides reliable results, particularly in the cooler atmospheres where water vapor plays a significant role in shaping the observed spectrum.

While the effective temperature and metallicity are mainly constrained in the method by the OH and Fe lines chosen, respectively,  selected sets of weak and strong CO and CN lines  help constrain the microturbulence (\micro). The carbon and nitrogen abundances are simultaneously derived from these CO and CN molecular lines. The lines used are provided in \citet{Nandakumar:2023}.

\subsubsection{Surface Gravity}

\begin{figure*}
  \includegraphics[width=\textwidth]{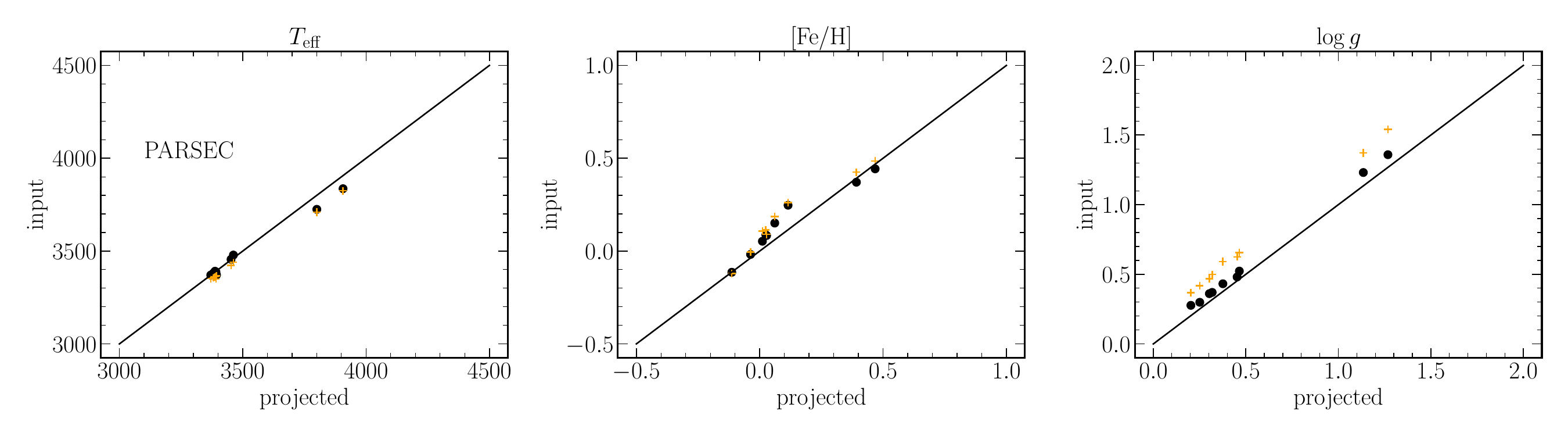}\\
  \includegraphics[width=\textwidth]{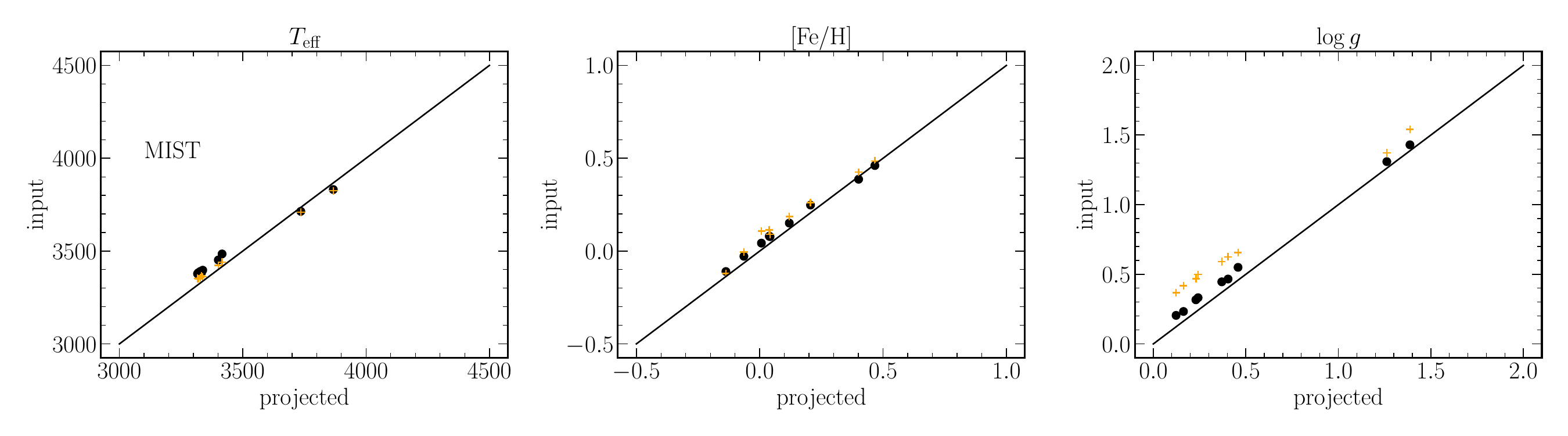}\\
  \includegraphics[width=\textwidth]{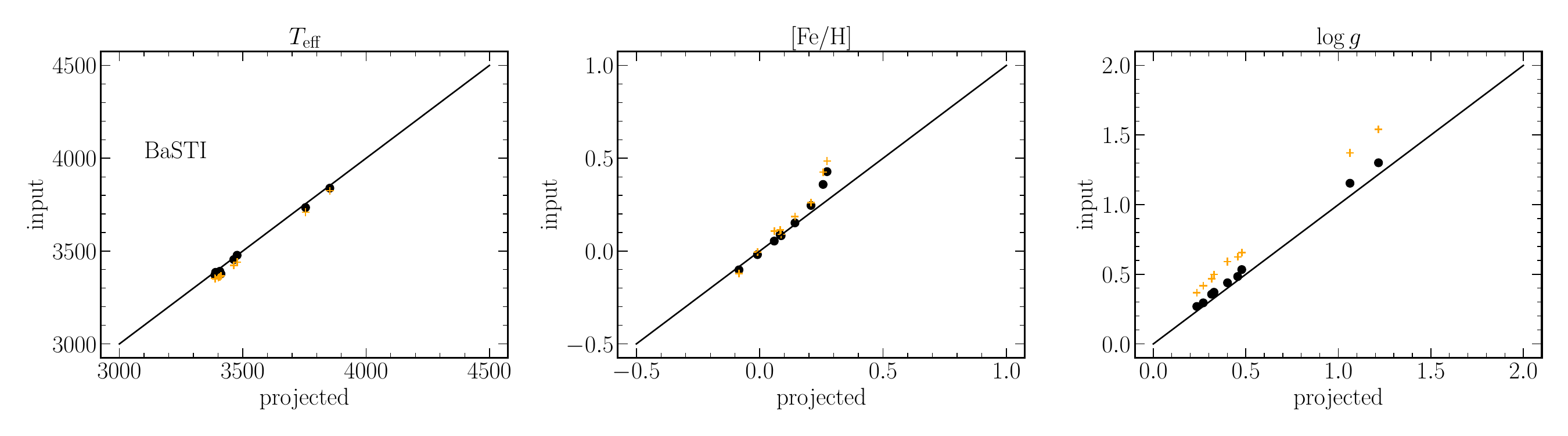}
  \caption{Input parameters versus isochrone-projected ones, for \teff\ (first column), [Fe/H] (second column) and \logg\ (third column) using the \citet{Kordopatis2023a} code. First row shows the projection on PARSEC isochrones, second on MIST, and third on BaSTI (solar scaled). Points far from the diagonal indicate that no isochrones are found close to the input values.  Yellow `+' signs are the atmospheric parameters from the first iteration (\logg\ obtained from the YY isochrones), whereas the black points are the results for a second (and final) projection, once new spectroscopic \teff\ and metallicities are determined taking into account the  projected \logg\, from the first iteration. One can notice that a very good convergence is reached in that case. } \label{fig:iso_projection}
\end{figure*}

The surface gravities of the stars were determined using Yonsei-Yale (YY) isochrones, based on the stars' effective temperatures and metallicities \citep{Demarque:2004}. The isochrones assume old stellar ages of 2–10 Gyr, which is appropriate for low-mass giants.   
To investigate the impact of using different sets of isochrones and removing the restriction to old isochrones, we employ the Bayesian isochrone projection code developed by \citet{Kordopatis2023a}.  Originally developed to determine stellar ages, masses and radii, this code also allows for the  estimation of the most likely atmospheric stellar parameters based on a set of isochrones. In this case, we will hence compare a set of input parameters (our derived \teff, \logg, and [Fe/H]) by projecting them onto different sets of isochrones. The sets considered are  PARSEC \citep{Bressan2012}, MESA \citep{Dotter2016} and BaSTI \citep{Hidalgo2018}. 
We ran the code by assuming uncertainties of $100$\,K, $0.2$\,dex, $0.1$\,dex, for \teff, \logg\ and [Fe/H], respectively, and without any prior on age versus metallicity. Isochrones were spaced with a log-age step of 0.05, and a metallicity step of 0.05\,dex. 
The results of the projections are displayed in Figure~\ref{fig:iso_projection} (orange `+' symbols) for the three sets of isochrones, with PARSEC, MESA (MIST), and BaSTI isochrones presented from top to bottom, respectively. What we find is that the best projection of all isochrone sets yield surface gravities that are close to and consistent with the original ones. This means partly that the exact choice of isochrone-set is not critical, and partly that the assumption of using an old-aged YY isochrone set works well. Isochrones of different ages do, indeed, lie close to each other on the giant branches. Our tests show that the spectroscopic \logg\, is $\sim 0.1$\,dex larger than the projected values, no matter the isochrones. Implementing the projected \logg\ as an additional  iteration in our stellar parameters determination has a small impact. The \teff\ of our stars decreases by 0 to 40 K, the metallicities increase by approximately 0.05 dex, and the microturbulence increases by around 0.05\,\kms.

In their analysis of \logg-sensitive line wings in near-infrared spectra, \citet{jess_master} demonstrated that for the stars studied in \citet{Nandakumar:2023}, the surface gravities determined spectroscopically in this way from high S/N spectra are within 
$\pm0.2$\,dex of those obtained using the isochrone method described in \citet{Nandakumar:2023}. Additionally, they found that for the same stars, the surface gravities derived from the Bayesian isochrone-fitting tool, PARAM \citep[based on MESA isochrones;][]{param1, param2, param3}, agree within $0.1$\,dex with those in \citet{Nandakumar:2023}, assuring that no significant systematic effects are present.

As an additional test of our method for determining surface gravity, we reference the work of \citet{sheshshayana2024}, who evaluated the performance of the same iterative stellar parameter determination method (\logg, from YY-isochrones) used in this study by applying it to six field M giant stars from the \textit{Kepler} database. They determined the stellar parameters for these stars from their high resolution (R $\sim$ 50,000) GIANO-B spectra and then compared the resulting \logg\, values with the asteroseismic \logg-values in the APOGEE and Kepler Asteroseismology Science Consortium (APOKASC-3) catalog. They find overall good agreement for all six stars with a minimum difference 
of $+0.01$ dex and a maximum difference of $+0.18$ dex between the isochrone-derived and asteroseismic \logg\, values (see their Table 4 for more details). These differences are consistent with the estimated uncertainty of $\pm$0.2 dex in \logg\, from this method.

\subsection{Spectral Lines and Stellar Abundances}
\label{sec:stellarabund}

For the determination of the stellar abundances of Mg, Si, and Ca, we thus base our line list on the studies by \citet{Nandakumar:2023, Nandakumar:24_21elements}. 
We have also added two additional silicon lines in the K band ($\lambda_\mathrm{air}=21368.7$ and $21874.2$\,\AA). In our analysis, we incorporate departure coefficients from non-local thermodynamic equilibrium (non-LTE) grids for the elements C, N, O, Mg, Si, Ca, and Fe \citep{amarsi:20,amarsi:22}. 
The line data for the CO, CN, OH, and H$_2$O lines were adopted from the line lists of \citet{li:2015}, \citet{brooke:2016},  \citet{sneden:2014}, and \citet{exomol_h2o}, respectively. The molecular lines are important since they might blend with and affect the abundance determination from many of the atomic spectral lines. We have scaled the finally derived abundances with respect to the solar abundance values from \citet{solar:sme}.

\subsection{Reference Sample}

\citet{Nandakumar:2023,Nandakumar:24_21elements} also provide a detailed abundance analysis of a comparison sample of 50 M giants in the solar neighborhood, observed using the same observational setup and analyzed with the same method, including the same stellar parameter scale and spectral lines, as employed in our study presented here. The primary advantage is the ability to directly and differentially compare our findings for the NSC stars with the abundance trends for the solar neighborhood sample, element by element and line by line. This approach minimizes systematic uncertainties, with the only difference between the analysis of the NSC sample and the solar neighborhood sample being the slightly lower S/N of the former. In the following sections, we adopt the terminology from \citet{Haywood:2013,Dimatteo:2016} 
by referring to the inner-disk sequence, when mentioning the thick disk and metal-rich thin disk (see Section \ref{sec:discussion}).

\section{Results}
\label{sec:results}

The effective temperatures for our NSC stars range from 3350 and 3850\,K, while those of the solar neighbourhood sample fall within the range of $3400\,\lesssim$\teff $\,\lesssim4000$\,K. The surface gravities, \logg, lie between 0.3 and 1.5 and the metallicities between $-0.1 \lesssim$\feh $\lesssim0.5$. Table~\ref{table:parameters} provides the final stellar parameters of the NSC stars, as well as their [C/Fe], [N/Fe], and assumed [O/Fe] abundances.

\subsection{$\alpha$-Element Trends}

\begin{figure*}
  \includegraphics[trim={1cm 0 1cm 0},clip,angle=-90,width=\textwidth]{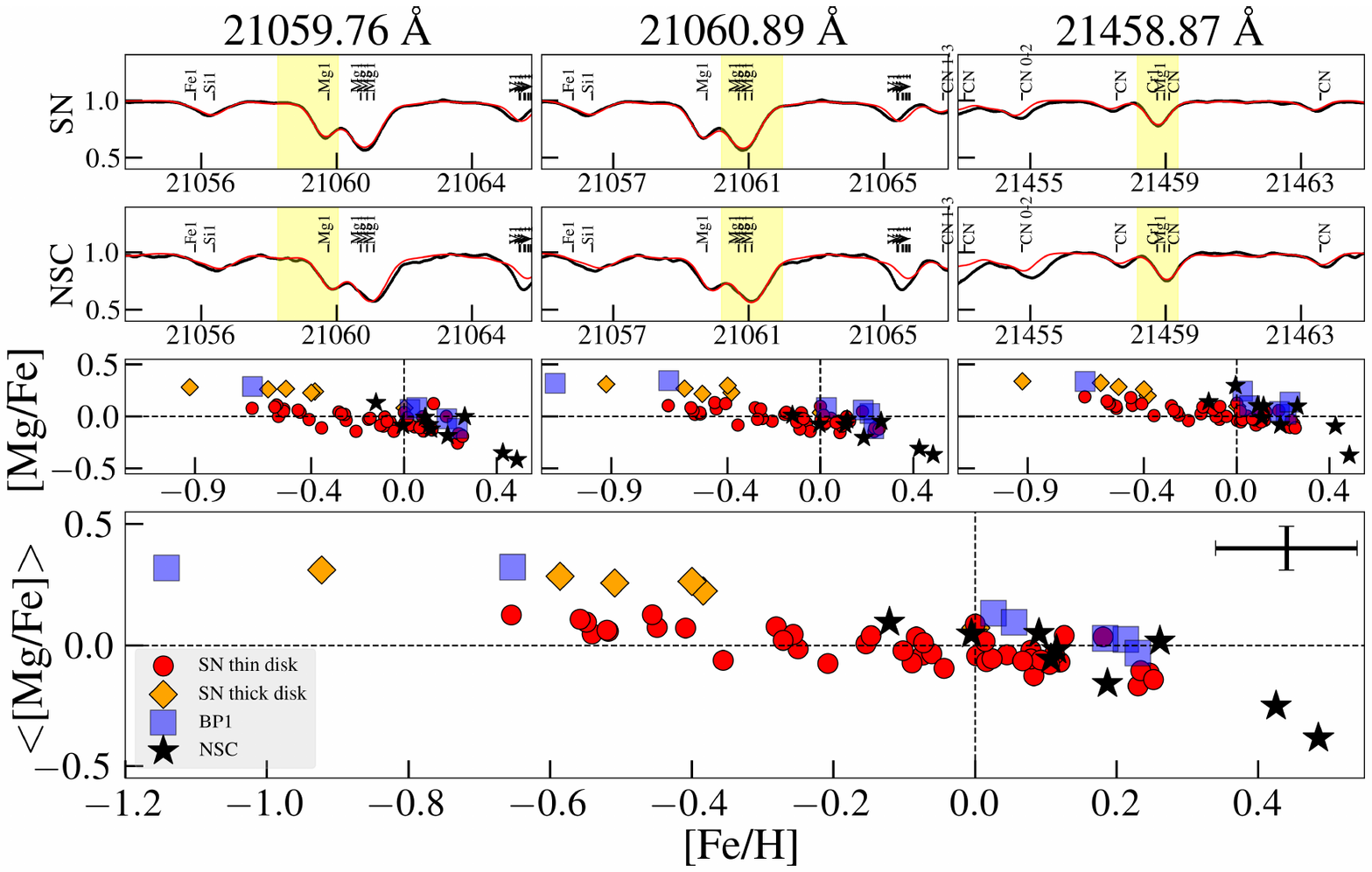}
  \caption{[Mg/Fe] versus [Fe/H] for different stellar populations. The Nuclear Star Cluster (NSC) stars are represented by black stars, the inner bulge stars from \citet{nandakumar:24} by blue squares, and the solar neighborhood thin-disk stars are depicted by red filled circles, while the thick-disk stars are shown as orange diamonds. In the upper two panels, the spectral lines used for the analysis are displayed, with a typical solar neighborhood star shown above and a typical NSC star (FK48) shown below. The trends derived from each individual spectral line are presented, along with the mean trend, which is displayed in the largest, bottom panel. }
  \label{fig:mg_trend}
\end{figure*}

\begin{figure*}
  \includegraphics[trim={3cm 0 3cm 0},clip,angle=-90,width=\textwidth]{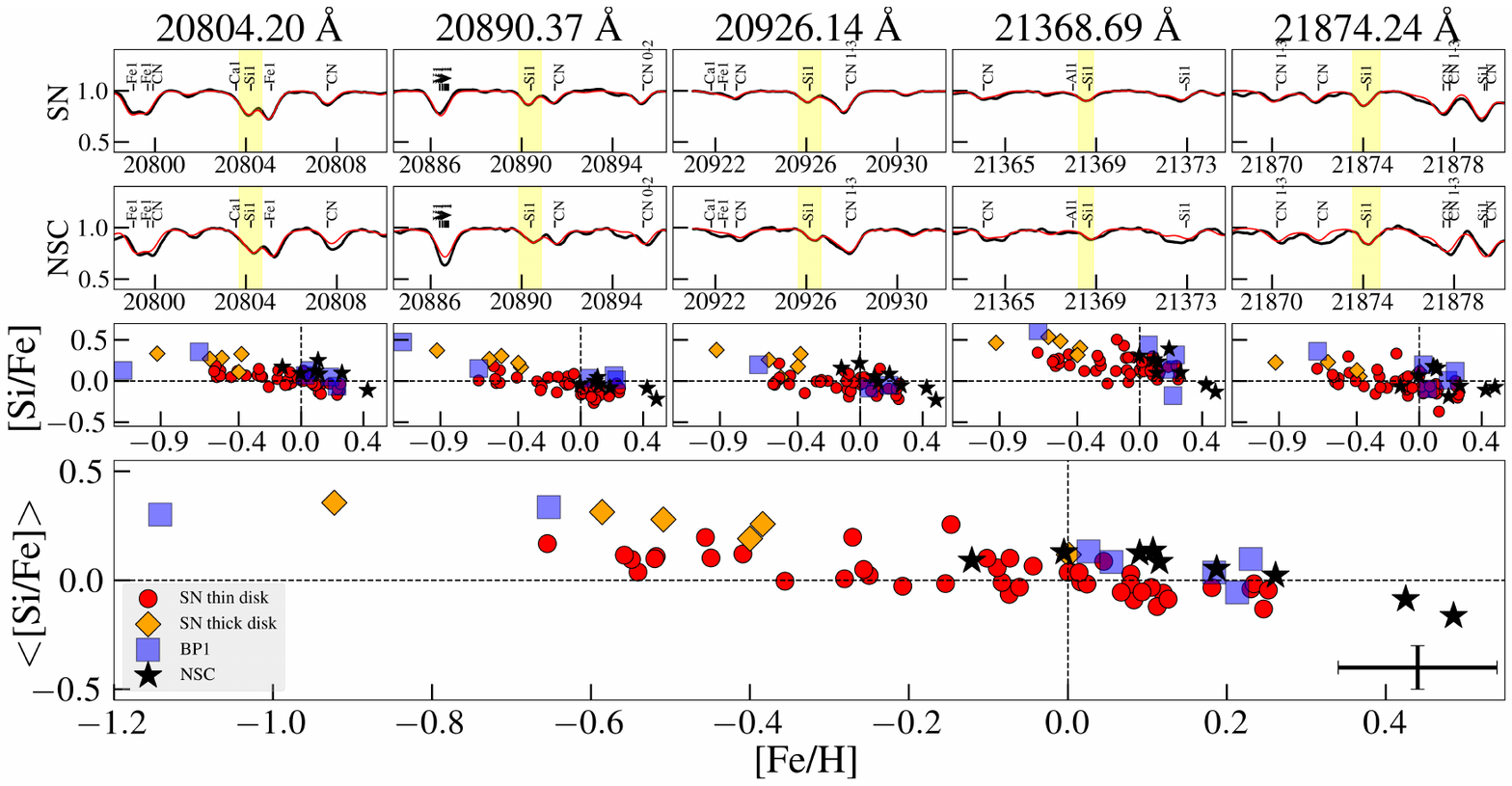}
  \caption{[Si/Fe] versus [Fe/H] for different stellar populations: NSC stars (black stars), inner bulge stars (blue squares), thin-disk stars (red circles), and thick-disk stars (orange diamonds). The upper panels display the spectral lines used, with a typical solar neighborhood star above and an NSC star (FK48) below. Trends from individual lines and the mean trend are shown in the bottom panel.}
  \label{fig:si_trend}
\end{figure*}

\begin{figure*}
  \includegraphics[trim={2cm 0 3cm 0},clip,angle=-90,width=\textwidth]{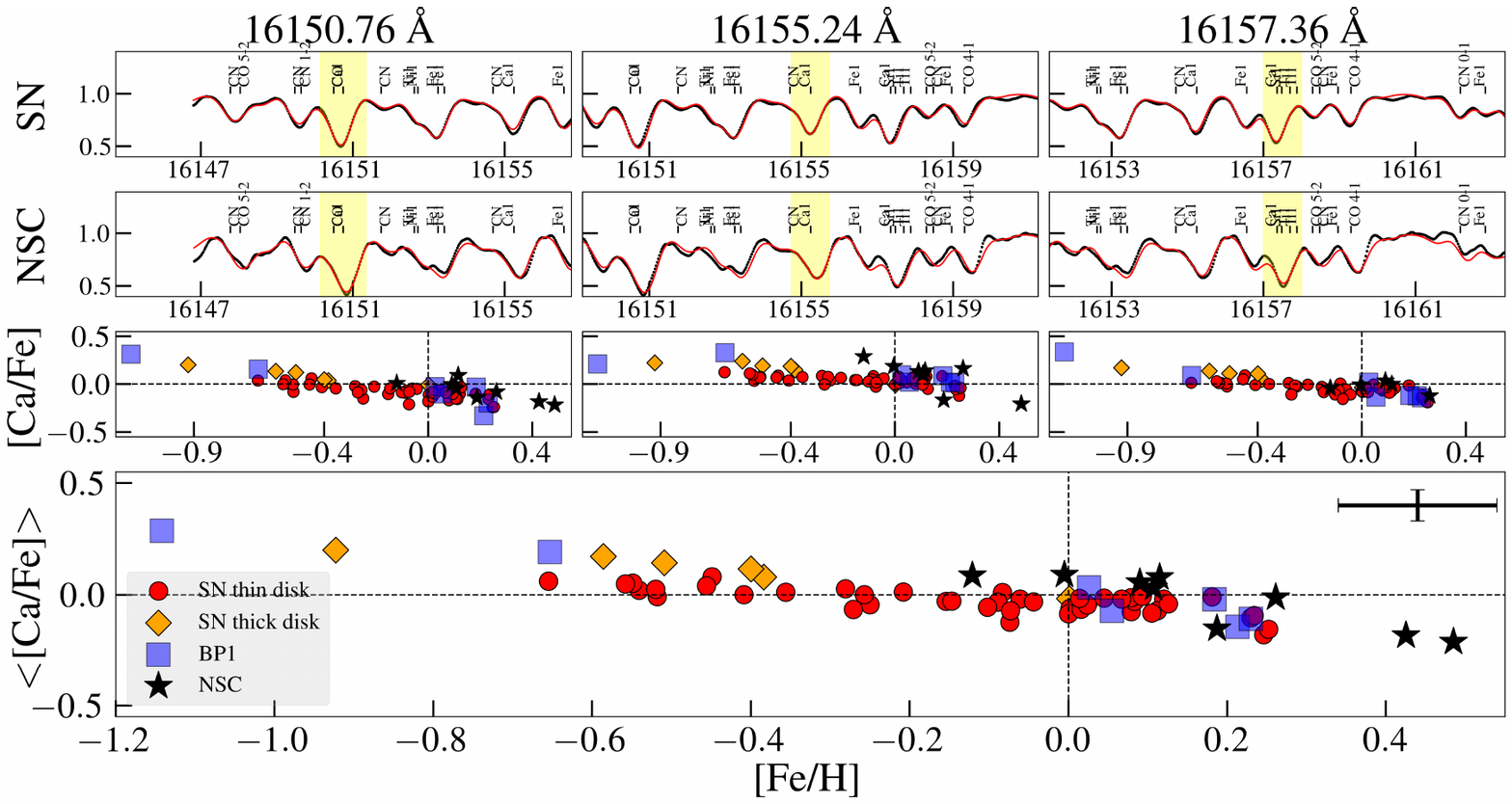}
  \caption{[Ca/Fe] versus [Fe/H] for different stellar populations: NSC stars (black stars), inner bulge stars (blue squares), thin-disk stars (red circles), and thick-disk stars (orange diamonds). The upper panels display the spectral lines used, with a typical solar neighborhood star above and an NSC star (FK48) below. Trends from individual three lines and the mean [Ca/Fe] trend versus metallicity  are shown in the bottom panels.}
  \label{fig:ca_trend}
\end{figure*}

Our results for the abundance trends of the $\alpha$ elements  as a function of metallicity  are presented in Figures \ref{fig:mg_trend} to \ref{fig:ca_trend}, with each figure corresponding to a different element. The trends for the stars in the NSC (black star symbols) are compared with those observed in the solar neighbourhood including low-$\alpha$ or thin-disk stars (in red circles)  and high-$\alpha$ or thick-disk stars (in orange diamonds). Additionally, the trends determined by \citet{nandakumar:24} for inner-bulge stars located 1 degree North of the Galactic Center are indicated by blue squares. These inner-bulge stars display a trend following the inner-disk sequence, aligning with the high-[$\alpha$/Fe] envelope of the metal-rich, thin-disk population in the solar vicinity, but having the same $\alpha$-enhancements as the solar vicinity thick-disk stars of the same [Fe/H] \citep[cf.][]{Haywood:2013}.

In the Figures, we display both the final mean trends and the trends derived on a line-by-line basis. This approach allows us to examine any systematic effects that may arise for different spectral lines.  The spectral lines used are based on the recommendations by \citet{Nandakumar:2023}, who identify lines that yield reliable results for high-quality spectra of M giants in the solar neighbourhood. For magnesium (Mg) and silicon (Si), three and five K-band lines were used, respectively. For calcium (Ca), three
H-band lines were used. The line wavelengths are also given in the Figures \ref{fig:mg_trend}-\ref{fig:ca_trend}. 

The systematic uncertainties are difficult to quantify; however, our differential analysis ensures that these uncertainties impact both samples in a similar way. The main random uncertainty in the determined elemental abundance ratios are due to the uncertainties in the stellar parameter determination.  We follow the determination of the uncertainties in the abundances as outlined in \citet{Nandakumar:24_21elements} and \citet{nandakumar:24} for the solar neighbourhood stars and the stars at ($l,b$)=(0$^{\circ}$,+1$^{\circ}$), respectively. In summary, we recalculated 50 abundance values for different sets of stellar parameters, randomly selected from normal distributions with  standard deviations representing typical uncertainties ($\pm$100 K in \teff, $\pm$0.2 dex in \logg, $\pm$0.1 dex in \feh, and $\pm$0.1 km s$^{-1}$ in $\xi_\mathrm{micro}$).  The abundance uncertainties range from 0.05 to 0.15, and are shown as error bars in the trend Figures   \ref{fig:mg_trend} - \ref{fig:ca_trend}. 

In the Figures \ref{fig:mg_trend} to \ref{fig:ca_trend}, we also provide examples of  typical line appearances from high S/N spectra of M giants in the solar neighborhood, as well as the line appearances from the typical NSC star FK48 (\teff$ = 3440$\,K, \logg$ = 0.6$, \feh $= + 0.12$, see Table\,\ref{table:parameters}).
These serve to illustrate typical line strengths that were used in the abundance analysis, as we aim to avoid lines that are too weak, which may disappear in the noise, as well as lines that are too strong, which may be significantly affected by the uncertain value of the microturbulence or formed in the outer, tenuous  regions of the stellar atmosphere where the structure is not well modeled.

 \begin{figure*}
  \includegraphics[trim={1cm 0 1cm 0},clip,angle=-90,width=\textwidth]{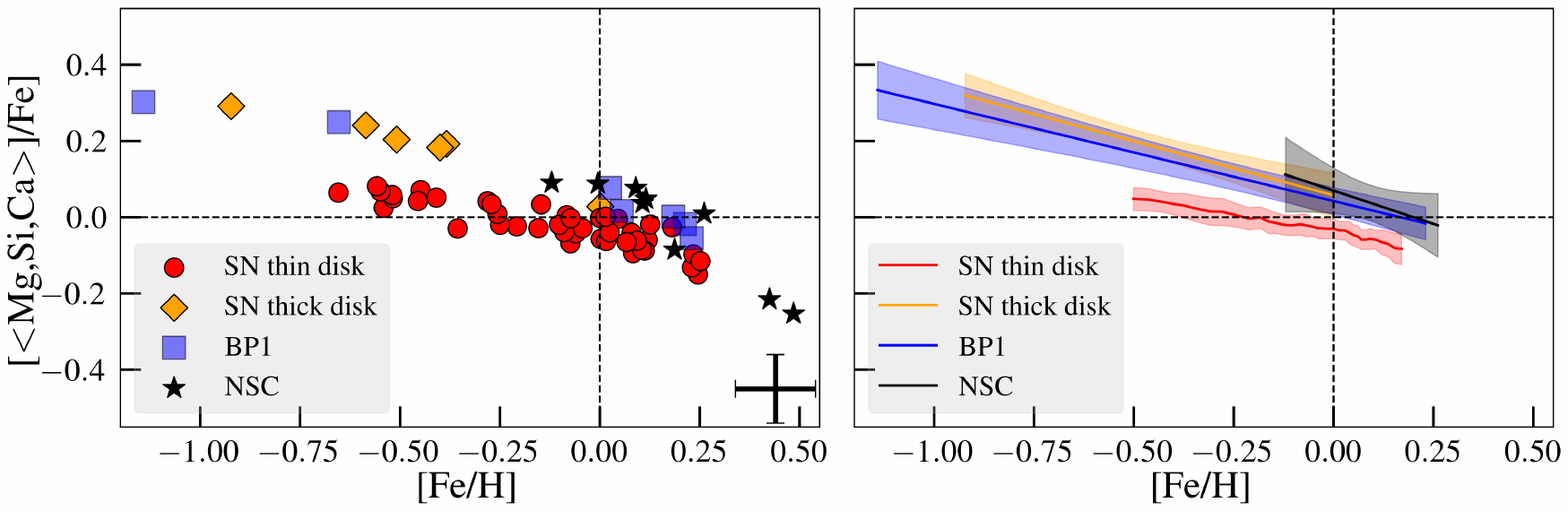}
  \caption{Straight mean abundance of the three $\alpha$ elements, Mg, Si, and Ca versus [Fe/H] for the NSC (black), inner-bulge stars located 1 degree North of the Galactic Center (blue), thick-disk (orange), and thin-disk stars (red). Left panel: Scatter plot of the mean $\alpha$ abundances. Right panel: Running mean for the solar neighbourhood trend and simple polynomial fits (solid line) to the mean $\alpha$ abundances and the respective standard deviations (respective colored band).  This shows a slight difference ($\sim$1-$\sigma$) between the trends of the solar neighbourhood and NSC populations (also thick disk and inner bulge). We excluded the two most metal rich stars in the NSC for the fit since there are no similar metallicity stars in the solar neighborhood stellar population.}
  \label{fig:alphamean_trend}
\end{figure*}

\subsection{Individual $\alpha$-Element Trends}

Overall, we observe clear enhancements in the $\alpha$-element abundances for the NSC stars,  consistent with 
the trend found for stars located 1 degree North of the Galactic Center, which we will refer to as 'inner-bulge stars' in the following. Below, we examine the individual $\alpha$-element trends in detail.

{\it Magnesium.} The [Mg/Fe]-trend follows the inner-bulge trend within uncertainties, as shown in  Figure \ref{fig:mg_trend}. It is clearly following the upper envelope of the metal-rich part of the thin-disk stars (red). The [Mg/Fe] ratios of the two metal-rich stars at \feh$\gtrsim 0.4$ are clearly subsolar, extending a downward trend. 

{\it Silicon.} Similarly, the [Si/Fe]-trend also follows that of the inner-bulge stars, see Figure \ref{fig:si_trend}. The trend exhibits less scatter, which can be attributed to the use of  five different spectral lines, all of which display quite a small scatter. The most metal-rich stars also show subsolar [Si/Fe] ratios.

{\it Calcium.} The mean [Ca/Fe]-trend is derived from three spectral lines, lying in  quite blended regions in the H band, see Figure \ref{fig:ca_trend}. The surrounding spectra are, however, well modelled and continuum regions can be defined nearby, which gives us confidence in the accuracy of the derived Ca abundances. While some variations are observed among the trends of the individual lines, there is a clear decreasing trend in the [Ca/Fe] ratio with increasing metallicity, following the upper envelope of the thin-disk stars. 

The H-band lines are expected to have a lower S/N than several potentially useful K-band lines. The K-band lines are beautiful, strong, and seemingly unblended. However, the abundances derived from them show a larger scatter in all data sets, with some abundance values systematically and significantly higher than expected from the H-band lines.  This discrepancy between H-band and K-band lines, increasing with metallicity, was also addressed by \citet{koch-hansen:22} in their near-IR spectroscopic study observing K giants toward the bulge with the IGRINS spectrometer. They identified such a discrepancy for Ca, Al, and Ti but not for all elements in their study. They explored potential causes for this difference, such as uncertainties in line strengths, difficulties in continuum definition, or the effects of non-LTE, but did not identify a definitive cause for the discrepancy. Until these differences in abundances are understood, especially for high metallicities, we choose to only use the H-band lines of calcium. Thus, our final mean Ca trend shows a decreasing trend with metallicity, similar to the inner-bulge stars, with the two most metal-rich stars showing subsolar [Ca/Fe] ratios.

{\it Mean $\alpha$-abundance trend.} In the left panel of Figure \ref{fig:alphamean_trend}, we have combined the trends into a single, general $\alpha$-abundance trend, despite recognizing  that the different $\alpha$ elements should exhibit slightly different trends \citep[see, e.g.,][]{Woosley:1995, Matteucci:2021}. The NSC [$\alpha$/Fe] trend shows a clear and steady decrease with increasing metallicity, with the two most metal-rich stars displaying subsolar ratios. In order to approximately quantify the difference in trends between the different stellar populations, we carried out a simple polynomial fit to the mean $\alpha$-abundance trends and estimated the standard deviations. In the right panel of Figure \ref{fig:alphamean_trend}, we show the fit and the standard deviations with solid lines and similar colored bands respectively. For the solar neighbourhood trend we show the running mean instead. For the NSC population, we do not include the two most metal rich stars since there are no similar metallicity stars in the solar neighborhood population. The trends of the inner-bulge and NSC stellar populations are clearly similar and overlap with each other. Hence we cannot reject the hypothesis that they are drawn from a similar population. The {\it mean} trends of the NSC and that of the inner-bulge lie slightly higher than that of  the solar neighborhood metal-rich, thin-disk stars by approximately 0.1 dex in abundance, which slightly exceeds 1$\sigma$ of the NSC population. The NSC and inner-bulge trends, however, follow the upper envelope of the thin-disk stars, consistent with the inner-disk sequence distribution.

\section{Discussion}
\label{sec:discussion}

A detailed chemical characterization of the stellar populations in the Galactic Center region is essential for unraveling its formation history and understanding the relationships between the different structures located within it. Here, we will compare the diagnostically interesting $\alpha$-element trends in the NSC directly with a comparison set observed in the solar neighbourhood. We will also conduct a differential and direct comparison with the inner-bulge population located at $1^\circ$ North of the Galactic Center, as investigated by \citet{nandakumar:24}, using the same methodology applied in this study. 

The trend of the [$\alpha$/Fe] abundance ratio versus the metallicity, [Fe/H], of a stellar population, is useful for determining  the star-formation history of that population \citep[see, for instance,][]{McWilliam:97,Matteucci:2021}. The reason is that the [$\alpha$/Fe] abundance ratio found in stars depend partly on the $\alpha$ abundances in the star-forming cloud, abundances that is formed on short time-scales from massive stars (expelled during the core-collapse supernova (CCSNe) explosion), and partly on the iron abundance, which is formed on longer time-scales in thermonuclear supernovae (SN Type Ia, SN Ia). The star-formation rate will determine the increase of the metallicity with time. Magnesium is predominantly produced during the hydrostatic burning stage over a massive stars’s lifetime. Si and Ca are mainly explosive $\alpha$-elements and are produced in the CCSNe explosion \citep[see, e.g.,][]{Woosley:1995,Matteucci:2021}. Whereas magnesium is nearly solely formed in massive stars, it should be noted that, for instance, silicon is produced both in CCSNe and in SN Ia in almost equal proportions 
\citep{Matteucci:20}.  

Any differences in $\alpha$-element trends between populations would reveal environmental influences on, for example, the star-formation histories and processes, gas-flow patterns, evolutionary timescales, and stellar evolutionary processes in the different populations \citep[see, for instance,][]{friske:23}. The $\alpha$-element trends and their scatter are also useful for determining the spatial (e.g. in situ or accreted) and temporal origins of the stars in that population \citep[see, for example,][]{vintergatan1,vintergatan3}, as well as properties of the interstellar medium, e.g. the gas depletion time in star forming clouds \citep[see e.g. figure 9 in][]{vintergatan2}.

As a background, the established $\alpha$-abundance trends of the bulge/bar appear similar to a thick disk extending to higher metallicities \citep{melendez:08,Ryde2010,Kordopatis2015,Rojas-Arriagada:2017,Zasowski2019,lomaeva:19,Nieuwmunster:2023}. 
This can be compared to the picture suggested in \citet{Dimatteo:2016} where the thick-disk (high $\alpha$)  stellar population in the the solar neighbourhood is referred to as the ‘inner disk sequence’; the metal-rich, thin-disk population that is joined by the ‘inner-disk sequence’ is referred to as the ‘inner thin disk’ and it is considered to be the same structure as the thick-disk or ‘inner-disk sequence’ \citep[see also][]{Haywood:2013}. A comparable scenario is observed in the cosmological zoom-simulation  {\small VINTERGATAN} \citep{vintergatan1,vintergatan2} of a Milky Way-like galaxy, where the chemically-defined thin disk stars in the solar neighborhood are shown to have formed in an outer disk. Furthermore, a general view is emerging according to which the dynamical bar becomes dominant for [Fe/H]$>-0.5$\,dex \citep{soto2007,ness2013}, with the metal-rich stars more concentrated to the Galactic plane \citep{johnson2020,johnson2022}. This metal-rich structure of the bulge is most likely of a secular formation origin from the early disk \citep[][]{shen:10,Ness2012,Rojas-Arriagada:2017,Rojas2020,Zoccali:2017}. The investigation by \citet{nandakumar:24} of the inner-bulge population located at $1^\circ$ N showed a similarity between the inner-bulge population and the inner-disk sequence, following the high-[$\alpha$/Fe] envelope of the solar vicinity metal-rich, thin-disk population.

While the majority of stars in the Galactic Center are generally  very old ($>8$\,Gyr), with those in the NSC being more than 10 Gyr old \citep{schodel:20}, steady gas inflows channeled by the Galactic bar have indeed triggered bursts of younger star formation over time. It is believed that as much as 15\% of the total star formation gave rise to an intermediate-aged ($\sim3$\,Gyr) population in the NSC \citep{lara:21}. In
contrast, \citet{chen:23} do not identify the old population, but a dominant, metal-rich
component (constituting over 90\% of the stellar mass) with an age of around 5 Gyrs. The star formation history of the bulge as a whole is also found to be complex with a range of ages found for metal rich stars (\feh\,$>$0). For instance, \cite{Bensby:2017} estimated 35$\%$ of metal rich stars to be younger than 8 Gyr while \cite{Joyce:2023} estimated it to be only 18$\%$ by reassessing the ages of a subset of the same stars with a different isochrone set. 

Our $\alpha$-element trends  as a function of metallicity in the NSC all decrease with increasing metallicity also at the highest metallicities, which is expected from Galactic Chemical Evolution models \citep[see, e.g.,][]{grisoni:17, Matteucci:2021}. We also observe that the trends for Mg, Si, and Ca (as well as the mean $\alpha$-element trend shown in Figure \ref{fig:alphamean_trend}) show similarities to those observed in the inner bulge located at {$(l,b) = (0,+1^\circ)$} \citep{nandakumar:24}. These trends also align with the broader characteristics of the bar population and the inner-disk sequence,  following the high-[$\alpha$/Fe] envelope of the metal-rich population in the solar vicinity. This suggests that the chemical pattern observed in the NSC stars is consistent with that seen in solar vicinity thick-disk stars at the same [Fe/H] values, exhibiting similar levels of $\alpha$-element enhancement. This is interesting because the thick-disk sequence, extending to its metal-rich limit, appears to be the only chemical sequence in both the inner disk and the Bulge (the so-called inner-disk sequence). Thus, our NSC stars appear to be located on top of the chemical sequence found in the inner Galaxy, on much larger spatial scales.

Since the $\alpha$-elements versus [Fe/H] patterns are so similar, this suggests that the star-formation history (SFH) of the NSC (at these metallicities) and of the thick disk, at the same metallicities, must also have been similar. Because of the existence of a strong link between the SFH of the thick disk and the Bulge \citep[see, e.g, ][]{haywood:18},  this suggests a surprisingly similar evolution of the star formation from a several-kpc scale (the thick disk), to a few-kpc scale (the Bulge) up to the innermost central regions of the Milky Way (the NSC).  

Thus, the similarities suggest that the NSC may resemble the inner bulge, which in \citet{matteucci:19} is described as being {\it predominantly} old with a
rapid star-formation history. This would allow us to rule out a dominant 5 Gyr
starburst in favour of an 8–10 Gyr burst. In fact, in \citet{Nieuwmunster:2023}, a chemical evolution model was proposed to reproduce the $\alpha$-element abundance trends in the inner-bulge within $|b|<2^\circ$,  
showing peaks
in the star-formation rate  concentrated to the first 4 Gyr of
the  evolution of the bulge (i.e. ages older than $\sim10$\,Gyr). It is not
necessarily possible, however, to completely rule out the presence of a
3 Gyr-old burst responsible for the formation of an intermediate-aged
population in the NSC \citep{schodel:20,lara:21}. If such a burst accounts for a relatively small fraction of the total
star-formation rate, it may only cause a minor change in the observed
$\alpha$-element abundance pattern. This effect could be obscured by
uncertainties, and can only be conclusively ruled out through a detailed
chemical evolution model, which will be the focus of future
investigations. 
Investigating the influence and magnitude of early gas inflows into the NSC region on the observed $\alpha$-element trends in this study would also be highly valuable. However, we emphasize that our sample size is small, limiting the constraints we can place on the star-formation history of the NSC. The upcoming MOONS survey \citep{MOONS2020} will significantly increase the sample size in the NSC, helping to provide a more comprehensive understanding of the complex star-formation history in this region.

The detailed spectroscopic analysis of more than 1000 FGK dwarf stars in the solar neighborhood using the high-resolution (R$\sim$115 000) spectra from the HARPS-GTO program have revealed a population of high $\alpha$ metal rich (h$\alpha$mr) stars \citep[][]{Adibekyan:2011, DelgadoMena:2019}. The h$\alpha$mr population have \feh\, $>$ -0.2 dex, enhanced [$\alpha$/Fe] ratios with respect to the thin disk stars and are found to be older ($>$6 Gyr) than the thin disk stars at similar metallicities \citep[see Fig. 5 in][]{DelgadoMena:2019}. A link between the h$\alpha$mr stars and stars in the inner Galactic disk or the bulge have been proposed by \cite{Adibekyan:2011}. We do not find similar stars in our solar neighborhood sample since they are rare and hence difficult to find among such small sample of stars. We, therefore, note that the absence of the h$\alpha$mr stars in our reference sample further accentuates the enhancement in alpha abundances for the NSC and inner bulge populations compared to the metal rich thin disk stars. Hence we reassert that our NSC and inner-bulge stars follow the high $\alpha$ envelope of the solar vicinity metal-rich population.

Although our sample size is small and we have not accounted for potential selection effects, the metallicity distribution of the NSC stars analyzed in this study corroborates the broad abundance range reported in the first paper of this series \citep{Rich:2017}.  Specifically, we confirm the presence of stars with metallicities up to \feh \,$=+0.5$. We note that we find few stars with subsolar metallicities in our sample, but red giants with lower metallicities tend to be warmer, and we have intentionally excluded these stars from our analysis due to spectroscopic constraints. The distribution of metallicities as a function of latitude and longitude is otherwise a strong diagnostics and could be compared to that of the local thick disk and with model prediction of the modulation of the metallicity distribution function (MDF) with latitude and longitude, as observed over the whole bulge extent \citep[see, for e.g.,][]{fragkuodi:18}.  

At supersolar metallicities, we observe a clear decreasing trend in $\alpha$-element abundances; however, we do not detect the elevated [Si/Fe] values reported in the second paper of this series by \citet{thorsbro:2020}. It is possible that these high-$\alpha$ stars belong to a distinct population within the NSC. \citet{thorsbro:2020} acknowledged that their conclusions were based on a small sample size and highlighted the need to verify the trend using other $\alpha$-elements. Unfortunately, we lack IGRINS spectra for the stars analyzed in that study, which would not only have provided us with the [Mg/Fe] and [Ca/Fe] trends versus [Fe/H], but also have enabled a differential analysis. Such an analysis would clarify the influence of analysis differences due to the lower resolution, limited K-band coverage, application of non-LTE corrections, or possible differing temperature scales, or the cooler temperatures of the stars in \citet{thorsbro:2020} (3300–3450 K) compared to our metal-rich stars (3700–3800 K).
It cannot be excluded that there might be  a complex high-metallicity environment with a spread in [Si/Fe] values.  \citet{thorsbro:23} further suggest a complex chemical evolution and enrichment in the NSC by finding a wide metallicity range of 1.7 dex among three $\sim$Gyr-old NSC stars. A systematic survey of a larger sample of NSC stars, which will be provided by the MOONS survey, is necessary to identify and evaluate potential substructures within the NSC.

\section{Conclusions}
\label{sec:conclusion}

This study is part of a series characterizing the chemical evolution of the Nuclear Star Cluster (NSC) of the Milky Way and its relation to the Galactic inner regions, with implications for understanding the formation history and evolution of the Milky Way's central stellar populations. A detailed chemical abundances analysis is needed for such studies.  However, the Galactic Center has been largely unexplored chemically due to significant optical extinction along the line-of-sight. High-quality, high-resolution, near-infrared spectroscopic data are necessary to achieve the precision and accuracy required for such analysis. The IGRINS spectrometer mounted on a 10-meter class telescope, fulfills these criteria, covering the full H and K bands in its observations. 

Comparing the chemical histories of different stellar populations is a powerful tool. The abundance trends of the $\alpha$ elements are a key diagnostic, offering insights into the star-formation rate and in-fall history. In this study, we have, therefore, determined detailed abundances and their trends with metallicity for the $\alpha$-elements Mg, Si, and Ca in 9 M-giants in the NSC, observed using IGRINS on the Gemini South telescope. We performed a robust membership analysis of the target stars to ensure that they are indeed NSC stars. These results are compared to derived abundances of M giants in the solar neighborhood and of a sample of M-giants 
located $1^{\circ}$ North of the Galactic Center (i.e. beyond the NSD) presented in \citet{nandakumar:24}. To ensure a strictly differential analysis and minimize systematic uncertainties, we employed the same instrument, wavelength range, spectral resolution, reduction pipelines, and analysis techniques for determining the stellar parameters and elemental abundances across all samples. All stars are of similar type and span the same parameter space in terms of fundamental stellar parameters.

We find enhanced [$\alpha$/Fe] ratios for the predominantly metal-rich stars in the NSC.  We see similar trends for all of the investigated $\alpha$ elements, namely Mg, Si, and Ca. These $\alpha$-element trends 
closely align with those in the inner bulge located at $+1^\circ$ N derived by \citet{nandakumar:24}, 
following the upper envelope of the metal-rich, thin-disk stars in the solar neighborhood, and following the inner-disk sequence. This pattern suggests a higher star-formation rate in the inner regions compared to that of the solar neighborhood. 
Our findings  suggest that the NSC population likely shares a similar evolutionary history with the inner-bulge population. Given that the inner-bulge population is presumed to have formed through a rapid and early star-formation history \citep{matteucci:19}, our data do not rule out the 10 Gyr-old star formation dominance in the NSC described by \citet{schodel:20}, but would not fit with the scenario of a more recent, dominant burst of star formation proposed by \citet{chen:23}. Detailed Galactic Chemical Evolution models are, however, needed to test the effects on the $\alpha$-element enhancements of different star-formation histories. The similarity also points to an in-situ formation of, at least, the metal-rich population in the NSC of the Milky Way \citep{neumayer:20}.

In a forthcoming paper, we will explore  abundance trends for additional classes of elements, similar to the analysis of inner-bulge stars by \citet{nandakumar:24}, and examine their respective evolutionary timescales. The nucleosynthetic channels traced by these elements will offer further insight into the relationship between the NSC and the inner-disk sequence of the Milky Way. If this connection is confirmed, it would suggest that the chemical properties of extragalactic NSCs in galaxies of a mass like the Milky Way could serve as valuable proxies for understanding the evolutionary processes of their host galaxies, in a manner explored first by Pagnini et al. (2024).

Further investigations of chemical abundances in the young population should be conducted, as they would provide insights into the recent chemical evolution. This approach was pioneered by \citet{cunha:07}, who demonstrated elevated $\alpha$-element abundances at high metallicities.\\

{\bf \noindent Acknowledgements}\\
We would like to thank Bengt Edvarsson, Thomas Masseron, and Henrik Jönsson for help with the EXOMOL water list of \citet{exomol_h2o}. N.R.\ acknowledge support from the Swedish Research Council (grant 2023-04744) and the Royal Physiographic Society in Lund through the Stiftelsen Walter Gyllenbergs, Märta och Erik Holmbergs, and Henry och Gerda Dunkers donations. G.N.\ acknowledges the support from the Wenner-Gren Foundations (UPD2020-0191 and UPD2022-0059) and the Royal Physiographic Society in Lund through the Stiftelsen Walter Gyllenbergs fond. G.N.\ also acknowledges the support received from the Royal Swedish Academy of Sciences (Vetenskapsakademiens stiftelser). O.A. acknowledges support from the Knut and Alice Wallenberg Foundation, the Swedish Research Council (grant 2019-04659), and the Swedish National Space Agency (SNSA Dnr 2023-00164). B.T.\ acknowledges the financial support from the Wenner-Gren Foundation (WGF2022-0041). A.M.A. acknowledges support from the Swedish Research Council (VR 2020-03940) and from the Crafoord Foundation via the Royal Swedish Academy of Sciences (CR 2024-0015). This work used The Immersion Grating Infrared Spectrometer (IGRINS) was developed under a collaboration between the University of Texas at Austin and the Korea Astronomy and Space Science Institute (KASI) with the financial support of the US National Science Foundation under grants AST-1229522, AST-1702267 and AST-1908892, McDonald Observatory of the University of Texas at Austin, the Korean GMT Project of KASI, the Mt. Cuba Astronomical Foundation and Gemini Observatory.
This work is based on observations obtained at the international Gemini Observatory, a program of NSF’s NOIRLab, which is managed by the Association of Universities for Research in Astronomy (AURA) under a cooperative agreement with the National Science Foundation on behalf of the Gemini Observatory partnership: the National Science Foundation (United States), National Research Council (Canada), Agencia Nacional de Investigaci\'{o}n y Desarrollo (Chile), Ministerio de Ciencia, Tecnolog\'{i}a e Innovaci\'{o}n (Argentina), Minist\'{e}rio da Ci\^{e}ncia, Tecnologia, Inova\c{c}\~{o}es e Comunica\c{c}\~{o}es (Brazil), and Korea Astronomy and Space Science Institute (Republic of Korea).
The following software and programming languages made this
research possible: TOPCAT (version 4.6; \citealt{topcat}); Python (version 3.8) and its packages ASTROPY (version 5.0; \citealt{astropy}), SCIPY \citep{scipy}, MATPLOTLIB \citep{matplotlib} and NUMPY \citep{numpy}.

\bibliographystyle{aasjournal}

\end{document}